\documentclass[english,dvips,12pt]{iopart}
\usepackage{babel}
\usepackage{iopams}
\usepackage{xcolor}
\usepackage{graphicx}
\usepackage{amsopn}
\usepackage{cite}
\usepackage{url}
\usepackage{array}
\usepackage{booktabs}

\renewcommand{\phi}{\varphi}

\usepackage{color}

\newcommand{\inrim}{INRIM}

\begin{document}
\bibliographystyle{unsrt}
\title[Analysis of QHE device connections]{\textbf{Matrix method analysis of quantum Hall effect device connections}}
\author{M Ortolano\textsuperscript{1,2} and L Callegaro\textsuperscript{2}}
\address{\textsuperscript{1} Dipartimento di Elettronica, Politecnico di Torino, Corso Duca degli Abruzzi, 24, 10129 Torino, Italy}
\address{\textsuperscript{2} \inrim\ --- Istituto Nazionale di Ricerca Metrologica,
Strada delle Cacce, 91, 10135 Torino, Italy}
\eads{\mailto{massimo.ortolano@polito.it}, \mailto{l.callegaro@inrim.it}}

\begin{abstract}
The modelling of electrical connections of single, or several, multiterminal quantum Hall effect (QHE) devices is relevant for electrical metrology: it is known, in fact, that certain particular connections allow i) the realization of multiples or fractions of the quantised resistance, or ii) the rejection of stray impedances, so that the configuration maintains the status of quantum standard. Ricketts-Kemeny and Delahaye equivalent circuits are known to be accurate models of the QHE: however, the numerical or analytical solution of electrical networks including these equivalent circuits can be difficult. In this paper, we introduce a method of analysis based on the representation  of a QHE device by means of the \emph{indefinite admittance matrix}: external connections are then represented with another matrix, easily written by inspection. Some examples, including the solution of double- and triple-series connections, are shown.
\end{abstract}
\maketitle

\section{Introduction}
National metrology institutes reproduce the unit of resistance, the ohm, by means of quantum Hall effect (QHE) devices. Present experiments (see~\cite{Jeckelmann:2005} for a recent review) are based on multiterminal (typically 8 terminals) semiconductor devices where a two-dimensional electron gas (2DEG) is created. At low temperature, and under appropriate values of magnetic flux density $\bi{B}$, the 2DEG resistance is quantised: $R_\textup{H} = R_\textup{K}/i$, where $R_\textup{K}=h/e^2$ is the von Klitzing constant and $i$ is the plateau index.

The accurate measurement of $R_\textup{H}$ has to reject any stray resistance (in dc) or impedance (in ac) which is added by the device contacts and the external wiring. Even though a simple four-terminal connection can be sufficient for dc measurements on a single device, the series or parallel connection of several devices~\cite{Delahaye:1993,Jeffery:1995}, the realization of single-chip quantum Hall arrays~\cite{Poirier:2009} and measurements in the ac regime~\cite{Schurr:2007} ask for more complex connection schemes, usually based on the \emph{double-} or \emph{triple-series} connections introduced by Delahaye~\cite{Delahaye:1993}. 

In single device double- and triple-series connections, only adjacent terminals are short-circuited together, and this yields a resistance value of $R_\textup{H}$. However, when non-adjacent terminals are short-circuited together, the resulting resistance value can be a multiple or a fraction of $R_\textup{H}$. Fang~\cite[figure 1(a)]{Fang:1984} experimentally investigated several connections achieving resistances values\footnote{Neglecting contact and wire resistances.} of $\displaystyle \frac{1}{3}R_\mathrm{H}, \frac{1}{2}R_\mathrm{H}, \frac{2}{3}R_\mathrm{H},  \frac{3}{2} R_\mathrm{H}, 2 R_\mathrm{H},  3 R_\mathrm{H}$ with the same 6-terminal device at a single plateau.

DC analytical modelling of double- and triple-series connections of single and twin devices was performed in~\cite{Delahaye:1993} and~\cite{Jeffery:1995}; in~\cite{Schurr:2007} the analysis was extended to the ac regime. The modelling process is based on the Ricketts-Kemeny model~\cite{Ricketts:1988} of the quantum Hall device, or on its derived ones~\cite{Delahaye:1993,Schurr:2007}. The above analyses were carried out by applying Kirchhoff's laws to the chosen model with constraints given by the external connections. Then, the resulting equations were solved to determine a particular four-terminal resistance (or impedance). Such analytical calculations can be tedious and error-prone, and numerical simulations can have problems of convergence~\cite{Sosso:1999}. Therefore, only a few connection schemes of known practical relevance have been analysed in full.

In this paper, we propose a general approach for modelling the electrical behaviour of multiterminal QHE devices with external connections. This method is based on the use of the so-called indefinite admittance matrix. Four-terminal resistances can then be determined from matrix equations depending on two matrices: one that represents the unconnected device (or devices); another that represents the external connections and that can be easily written by inspection. 

\section{The indefinite admittance matrix and its properties}
\label{sec:iam}
\begin{figure}
  \centering
  \includegraphics[clip=]{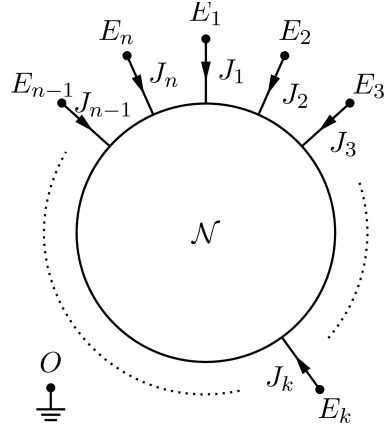}
  \caption{Multiterminal element representation with reference polarities and directions: terminal voltages are measured with respect to the arbitrary reference point $O$; terminal currents flowing into $\mathcal{N}$ are considered positive.}\label{fig:device_representation}
\end{figure}

Consider the $n$-terminal element $\mathcal{N}$ of figure~\ref{fig:device_representation} and let the terminal voltages\footnote{Throughout this work, voltages and currents are to be understood as voltage and current phasors.} $E_1,\ldots,E_n$ be measured with respect to an arbitrary reference point $O$ (datum node). We make the following assumptions:
\begin{enumerate}
\item[\textbf{A1}] $\mathcal{N}$ is linear and time-invariant.
\item[\textbf{A2}] Whenever $\mathcal{N}$ is connected to an external network, the complete circuit is uniquely solvable.
\item[\textbf{A3}] The sum of the terminal currents $J_1,\ldots,J_n$ is identically zero, i.e.\ $\sum_k J_k = 0$.
\item[\textbf{A4}] The terminal currents are invariant under a change of the reference point, i.e.\ under the transformation $E_k\mapsto E_k + E_0$.
\end{enumerate}

By means of assumptions \textbf{A1} and \textbf{A2}, when $\mathcal{N}$ is connected to an external network, we can write the relations between its terminal currents and voltages as
\begin{equation}\label{eq:terminal_currents}
J_k = \sum_{l=0}^n \bar{y}_{kl}E_l,\qquad k=1,\ldots, n,
\end{equation}
where the admittance coefficients $\bar{y}_{kl}$ are complex quantities defined by the equation
\begin{equation}
\bar{y}_{kl} = \left.\frac{J_k}{E_l}\right|_{E_j=0,j\neq l}.
\end{equation}
The set of equations~\eref{eq:terminal_currents} can be put in matrix form as
\begin{equation}
\bi{J} = \bar{\bi{Y}}\bi{E},
\end{equation}
where $\bi{J}$ and $\bi{E}$ are the column vectors $(J_1,\ldots,J_n)^\textup{T}$ and $(E_1,\ldots,E_n)^\textup{T}$ (T denotes the transpose operation), and $\bar{\bi{Y}}$ is the matrix $(\bar{y}_{kl})_{n\times n}$. This matrix, which has been called by Shekel~\cite{Shekel:1954} the \emph{indefinite admittance matrix}, has the following properties~\cite{Shekel:1954,Zadeh:1957,Sharpe:1960,Chua:1987}:
\begin{enumerate}
\item[\textbf{P1}] The sum of the elements of each column is identically zero, i.e.\ $\sum_k \bar{y}_{kl} = 0$ for any $l$. This property is a direct consequence of assumption \textbf{A3}.
\item[\textbf{P2}] The sum of the elements of each row is identically zero, i.e.\ $\sum_l \bar{y}_{kl} = 0$ for any $k$. This property is a direct consequence of assumption \textbf{A4}. Indeed, these first two properties imply that $\bar{\bi{Y}}$ is a singular matrix.
\item[\textbf{P3}] If its $r$th terminal is grounded, $\mathcal{N}$ can be regarded as an $(n-1)$-port with port voltages $E_1,\ldots,E_{r-1},E_{r+1},\ldots,E_n$. In this case, the non-singular $(n-1)\times (n-1)$ \emph{short-circuit admittance matrix} $\bi{Y}=(y_{kl})$ of the $(n-1)$-port can be obtained from $\bar{\bi{Y}}$ by deleting the $r$th row and column. Moreover, since $\bi{Y}$ is non-singular, it can be used to determine the $(n-1)\times (n-1)$ \emph{open-circuit impedance matrix} $\bi{Z}=(z_{kl})=\bi{Y}^{-1}$ of the $(n-1)$-port~\cite{Chua:1987}.
\end{enumerate} 

\begin{figure}
  \centering
  \includegraphics[clip=]{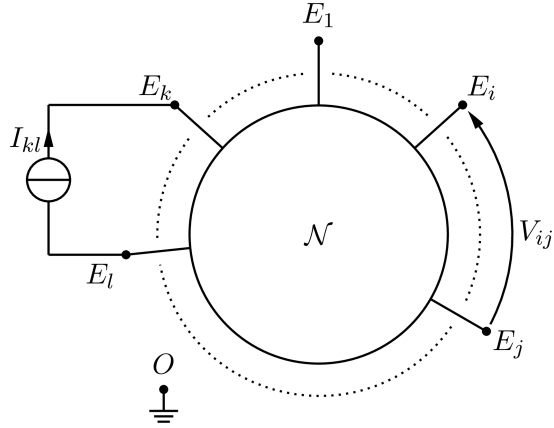}
  \caption{Circuit for the definition of four-terminal impedances.}\label{fig:impedances}
\end{figure}

In the application of QHE devices to resistance and impedance metrology, one is usually concerned with the determination of four-terminal transfer resistances and impedances. With reference to figure~\ref{fig:impedances}, we define the four-terminal impedance $Z_{ij,kl}$ as the ratio of the open-circuit voltage $V_{ij} = E_i-E_j$ to the current $I_{kl}$ flowing into terminal $k$ and out of terminal $l$, when all other terminals are open-circuited:  
\begin{equation}
Z_{ij,kl} = \left.\frac{V_{ij}}{I_{kl}}\right|_{I_m=0; m\neq k,l}.
\end{equation}
Throughout this work, impedances of the form $Z_{il,kl}$, i.e. with one terminal in common, are referred to as three-terminal impedances, and that of the form $Z_{kl,kl}$, i.e. with two terminals in common, are referred to as two-terminal impedances.

\begin{figure}
  \centering
  \includegraphics[clip=]{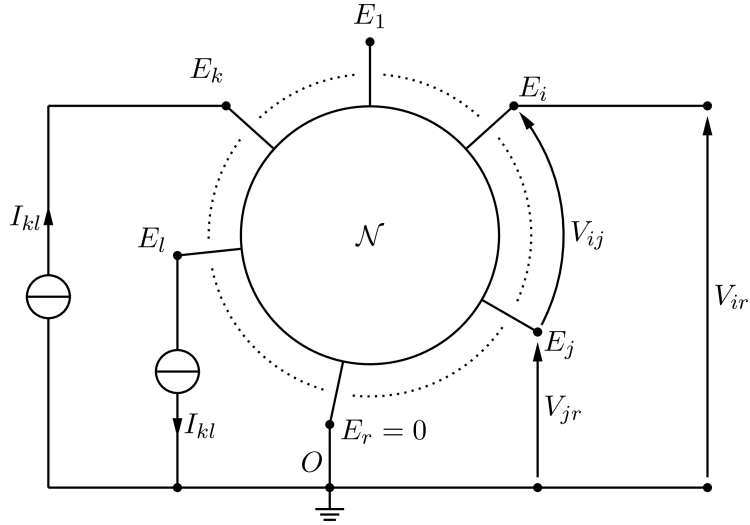}
  \caption{The $n$-terminal can be turned into an $(n-1)$-port by grounding one terminal (terminal $r$ in the figure).}\label{fig:impedances_grounded}
\end{figure}

One way of determining $Z_{ij,kl}$ from $\bar{\bi{Y}}$ makes use of property~\textbf{P3} above and is based on the fact that one can ``transform'' the $n$-terminal $\mathcal{N}$ into an $(n-1)$-port by grounding one terminal, as shown in figure~\ref{fig:impedances_grounded} for the $r$th terminal. In this case, the terminal pairs $kr$ and $lr$ can be considered as input ports, respectively driven by the currents $I_{kl}$ and $-I_{kl}$, while the terminal pairs $ir$ and $jr$ can be considered as output ports with port voltages $V_{ir}$ and $V_{jr}$, so that the ratio $V_{ij}/I_{kl} = (V_{ir}-V_{jr})/I_{kl}$ can be easily determined from the open-circuit impedance matrix $\bi{Z}$ obtained from the application of property~\textbf{P3}. This method is described in detail in section~\ref{sec:ideal_device_connections} by means of examples. A more straightforward, albeit more cumbersome, method of determining $Z_{ij,kl}$ from $\bar{\bi{Y}}$ is described in~\cite{Sharpe:1960}.

\section{The ideal QHE device}
\begin{figure}
  \centering
  \includegraphics[clip=]{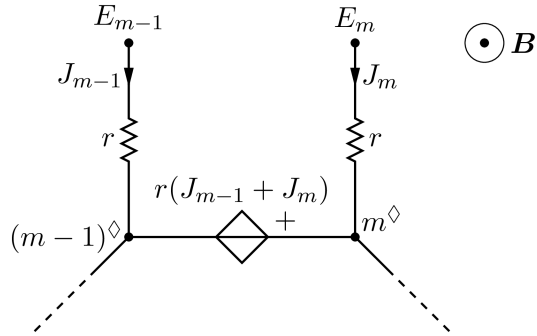}
  \caption{Ideal QHE device ring-array DC equivalent circuit (see text for explanation); $r = R_\textup{H}/2$.}\label{fig:device_model_ideal}
\end{figure}

Figure~\ref{fig:device_model_ideal} shows the portion between terminals $m-1$ and $m$ of the ring-array DC equivalent circuit of an \emph{ideal} $n$-terminal QHE device, as proposed by Delahaye~\cite{Delahaye:1993}. The magnetic flux density $\bi{B}$ points out of the page: reversing $\bi{B}$, reverses the sign of the current-controlled voltage sources. Applying Kirchhoff's voltage law (KVL) to nodes $m-1$, $(m-1)^\Diamond$, $m^\Diamond$ and $m$ yields 
\begin{equation}
J_m = G_\textup{H}(E_m-E_{m-1}),\qquad m=1,\ldots, n, 
\end{equation}
where $G_\textup{H} = 1/(2r) = 1/R_\textup{H}$ and it is assumed that $E_0 \equiv E_n$. The above set of equations directly determines the indefinite admittance matrix $\bar{\bi{Y}}_\textup{i}$ of the ideal QHE device: 
\begin{equation}\label{eq:iam_ideal_device}
\bar{\bi{Y}}_\textup{i} = G_\textup{H}\left(\begin{array}{rrrrrr}
1 &0 &\cdots  &0 &-1 \\
-1 &1 &0 & &0 \\
\vdots &-1 &1 &\ddots &\vdots \\
0 & & \ddots &\ddots &0   \\
0 &0 &\cdots &-1 &1
\end{array}\right),
\end{equation}
which is an $n\times n$ circulant matrix.

\section{Examples: single-device connections}
\label{sec:ideal_device_connections}
As was pointed out by Fang~\cite{Fang:1984} and Delahaye~\cite{Delahaye:1993}, multiterminal QHE devices can be connected in different ways to obtain multiples and fractions of the quantised Hall resistance. In this section, we describe how to analyse single-device connections by means of the indefinite admittance matrix.

\begin{figure}
  \centering
  \begin{minipage}[b]{0.2\textwidth}
  	\centering
  	\raisebox{12.03bp}{\includegraphics[clip=]{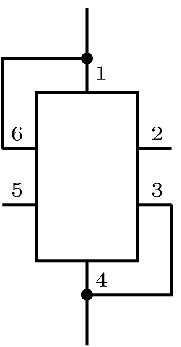}}
  \end{minipage}
  \begin{minipage}[b]{0.3\textwidth}
  	\centering
  	\raisebox{12.03bp}{\includegraphics[clip=]{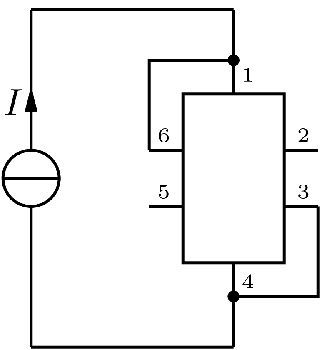}}
  \end{minipage} 
  \begin{minipage}[b]{0.3\textwidth}
  	\centering
  	\includegraphics[clip=]{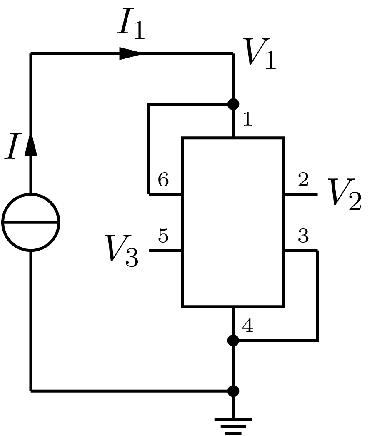}
  \end{minipage} 
  \\
  \begin{minipage}[t]{0.2\textwidth}
    \centering (a)  
  \end{minipage}
  \begin{minipage}[t]{0.3\textwidth}
  	\centering (b)
  \end{minipage} 
  \begin{minipage}[t]{0.3\textwidth}
  	\centering (c)
  \end{minipage} 
  \caption{Example of QHE device double series connection (see text for explanation).}\label{fig:single_device_1}
\end{figure}

Consider the simple example of figure~\ref{fig:single_device_1}(a), where a 6-terminal device is employed in a typical double-series connection with terminals 1,6 and 3,4 short-circuited. Suppose that the two-terminal impedance $Z_{14,14}$ is to be determined: this means, as shown in figure~\ref{fig:single_device_1}(b), that one has to inject a current $I$ between terminals 1 and 4 and to measure the voltage $E_1-E_4$ to obtain $Z_{14,14} = (E_1-E_4)/I$. Now, in order to determine $Z_{14,14}$, proceed as follow (see~\cite{Shekel:1954} for details).

\begin{enumerate}
\item Ground one terminal. In the above example, for simplicity, we have chosen to ground terminal 4, as shown in figure~\ref{fig:single_device_1}(c).
\item Choose a set $V_1,\ldots,V_{n'}$ of independent port voltages ($V_1, V_2, V_3$ in figure~\ref{fig:single_device_1}(c)). In general, since short-circuiting terminals constraints the terminal voltages, it is $n' < n-1$. 
\item Write down the relations between the terminal voltages and the port voltages: since these relations are linear, they can be written in matrix form as
\begin{equation} 
 \bi{E} = \bi{A}\bi{V},
\end{equation}
where, in general, $\bi{A}$ is an $n\times n'$ matrix and $\bi{V} = (V_1,\ldots,V_{n'})^\textup{T}$. For the circuit of figure~\ref{fig:single_device_1}(c),
\begin{equation}
\begin{array}{ll}
E_1 = E_6 = V_1, &\textup{(terminals 1 and 6 are short-circuited)} \\
E_2 = V_2, \\
E_3 = E_4 = 0, &\textup{(terminals 3 and 4 are grounded)} \\
E_5 = V_3,
\end{array}
\end{equation}
and therefore
\begin{equation}
\bi{A} = \left(\begin{array}{ccc}
1 &0 &0 \\
0 &1 &0 \\
0 &0 &0 \\
0 &0 &0 \\
0 &0 &1 \\
1 &0 &0
\end{array}\right).
\end{equation}
The corresponding port current vector $\bi{I} = (I_1,\ldots,I_{n'})^\textup{T}$ is related to the terminal current vector $\bi{J}$ by the equation 
\begin{equation}
 \bi{I} = \bi{A}^\textup{T}\bi{J},
\end{equation} 
so that the power flowing into the device is invariant, i.e.\ $\bi{V}^\textup{T}\bi{I} = \bi{E}^\textup{T}\bi{J}$.
\item Determine the $n'\times n'$ admittance matrix 
\begin{equation}\label{eq:Y}
\bi{Y} = \bi{A}^\textup{T}\bar{\bi{Y}}\bi{A}
\end{equation}
which relates the port current vector to the port voltage vector, $\bi{I} = \bi{Y}\bi{V}$. Since at least one terminal is grounded, by virtue of \textbf{P3}, $\bi{Y}$ is a non-singular short-circuit admittance matrix whose inverse is the open-circuit impedance matrix $\bi{Z} = \bi{Y}^{-1}$. For the circuit of figure~\ref{fig:single_device_1}(c), with $\bar{\bi{Y}} = \bar{\bi{Y}}_\textup{i}$, one obtains
\begin{equation}
\bi{Y} = G_\textup{H}\left(\begin{array}{ccc}
1 &0 &-1 \\ 
-1 &1 &0 \\ 
0 &0 &1
\end{array}\right)
\end{equation} 
and
\begin{equation}
\bi{Z} = R_\textup{H}\left(\begin{array}{ccc}
1 &0 &1 \\ 
1 &1 &1 \\
0 &0 &1
\end{array}\right).
\end{equation}
\item Identify the input and the output ports. In the example of figure~\ref{fig:single_device_1}(c), there are only one input port and one output port, and both coincide with port 1. Thus,
\begin{equation}
Z_{14,14} = \frac{E_1-E_4}{I} = \frac{V_1}{I_1} = z_{11} = R_\textup{H}.
\end{equation}
\end{enumerate} 

\begin{figure}
  \centering
  \begin{minipage}[b]{0.3\textwidth}
  	\centering
  	\raisebox{12.03bp}{\includegraphics[clip=]{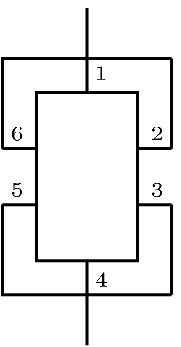}}
  \end{minipage}
  \begin{minipage}[b]{0.3\textwidth}
  	\centering
  	\includegraphics[clip=]{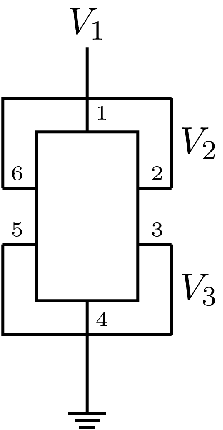}
  \end{minipage} 
  \\
  \begin{minipage}[t]{0.3\textwidth}
    \centering (a)  
  \end{minipage}
  \begin{minipage}[t]{0.3\textwidth}
  	\centering (b)
  \end{minipage} 
  \caption{Example of single-device connection.}\label{fig:single_device_2}
\end{figure}

A second, more significant, example is shown in figure~\ref{fig:single_device_2}(a). This is one of the connections experimentally analysed by Fang and Stiles~\cite[figure 1(a)]{Fang:1984}. In this case, suppose that the two-terminal impedance $Z_{14,14}$ and the four-terminal impedance $Z_{23,14}$ are to be determined. In particular, for $Z_{14,14}$, Fang and Stiles measured the value $Z_{14,14}\approx 3R_\textup{H}$ (within few parts in $10^4$).  

In figure~\ref{fig:single_device_2}(b), following the method described above, we have grounded one of the terminals (terminal 4, again, to have one input port only) and we have defined $V_1, V_2, V_3$ as port voltages. Thus
\begin{equation}
\begin{array}{ll}
E_1 = V_1, \\
E_2 = E_6 = V_2, &\textup{(terminals 2 and 6 are short-circuited)} \\
E_3 = E_5 = V_3, &\textup{(terminals 3 and 5 are short-circuited)} \\
E_4 = 0, &\textup{(terminal 4 is grounded)}
\end{array}
\end{equation}
and 
\begin{equation}
\bi{A} = \left(\begin{array}{ccc}
1 &0 &0 \\
0 &1 &0 \\
0 &0 &1 \\
0 &0 &0 \\
0 &0 &1 \\
0 &1 &0
\end{array}\right).
\end{equation}   

The short-circuit admittance matrix and the open-circuit impedance matrix are respectively
\begin{equation}
\bi{Y} = G_\textup{H}\left(\begin{array}{ccc}
1 &-1 &0 \\ 
-1 &2 &-1 \\ 
0 &-1 &2
\end{array}\right)
\end{equation} 
and
\begin{equation}
\bi{Z} = R_\textup{H}\left(\begin{array}{ccc}
3 &2 &1 \\ 
2 &2 &1 \\
1 &1 &1
\end{array}\right).
\end{equation}
Then, to determine $Z_{14,14}$: i) inject a current between terminals 1 and 4, i.e.\ \emph{in} port 1, which is considered as an input port; and ii) determine the voltage across terminals 3 and 4, i.e.\ \emph{across} port 1, which is also considered as an output port. Hence  
\begin{equation}
Z_{14,14} = \frac{V_1}{I_1} = z_{11} = 3R_\textup{H},
\end{equation}  
which is in agreement with the value given in~\cite{Fang:1984}.
Moreover, to determine $Z_{23,14}$: i) inject a current between terminals 1 and 4, i.e.\ \emph{in} port 1, which is considered as an input port; and ii) determine the voltage across terminals 2 and 3, i.e.\ \emph{across} port 2 and 3, which are considered as  output ports. Hence
\begin{equation}
Z_{23,14} = \frac{V_2-V_3}{I_1} = \frac{V_2}{I_1}-\frac{V_3}{I_1} = z_{21}-z_{31} = R_\textup{H}.
\end{equation}

\section{Modelling the contact resistances}
\begin{figure}
  \centering
  \includegraphics[clip=]{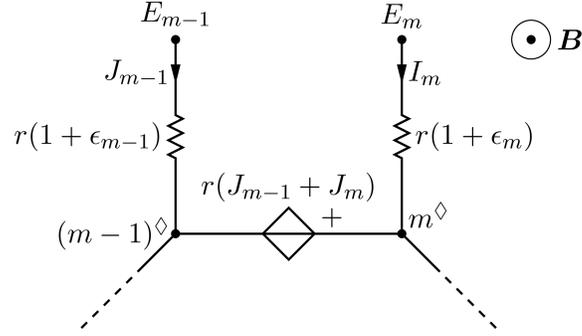}
  \caption{Non ideal QHE device DC equivalent circuit: contact resistances $\epsilon_m r$ ($m=1,\ldots,n$) have been added in series to each terminal.}\label{fig:device_model_nonideal}
\end{figure}

In the equivalent circuit of figure~\ref{fig:device_model_nonideal}, a contact resistance $\epsilon_m r$ has been added in series to each terminal. Applying KVL to nodes $m-1$, $(m-1)^\Diamond$, $m^\Diamond$ and $m$ yields
\begin{equation}
-\frac{\epsilon_{m-1}}{2}J_{m-1}+\left(1+\frac{\epsilon_m}{2}\right)J_m = G_\textup{H}(E_m-E_{m-1}),\qquad m=1,\ldots, n.
\end{equation}
In matrix form, the above set of equations can be written as
\begin{equation}
\bi{D}\bi{J} = \bar{\bi{Y}}_\textup{i}\bi{E}
\end{equation} 
or
\begin{equation}\label{eq:J_nonideal_device}
\bi{J} = \bi{D}^{-1}\bar{\bi{Y}}_\textup{i}\bi{E},
\end{equation}
where $\bar{\bi{Y}}_\textup{i}$ is given by~\eref{eq:iam_ideal_device} and
\begin{equation}\label{eq:D}
\bi{D} = \left(\begin{array}{cccccc}
1+\epsilon_1/2 &0 &\cdots  &0 &-\epsilon_n/2 \\
-\epsilon_1/2 &1+\epsilon_2/2 &0 & &0 \\
\vdots &-\epsilon_2/2 &1+\epsilon_3/2 &\ddots &\vdots \\
0 & &\ddots &\ddots &0   \\
0 &0 &\cdots &-\epsilon_{n-1}/2 &1+\epsilon_n/2
\end{array}\right).
\end{equation}
Because of~\eref{eq:J_nonideal_device}, the indefinite admittance matrix for the circuit of figure~\ref{fig:device_model_nonideal} is
\begin{equation}\label{eq:iam_nonideal_device}
\bar{\bi{Y}} = \bi{D}^{-1}\bar{\bi{Y}}_\textup{i}.
\end{equation}   

Actually, in analytical calculations, $\bi{D}^{-1}$ need not to be evaluated in full. In fact, since $\bi{D}$ is typically very close to the identity matrix (cf.\ equation~\eref{eq:D}), a Taylor series expansion of $\bi{D}^{-1}$ up to the appropriate order in $\epsilon_1,\ldots,\epsilon_n$ might be sufficient in most cases. 

\section{Examples: single device connections}
In this section we analyse the effect of contact resistances on several examples of single device connections (table~\ref{tab:single_device_connections}, first column). These examples, which were already analysed using other methods by Delahaye~\cite{Delahaye:1993} and by Jeffery \emph{et al.}~\cite{Jeffery:1995}, have been chosen for their practical interest. For each connection, the second column of table~\ref{tab:single_device_connections} shows a suitable choice of port voltages and the third column shows the corresponding matrix $\bi{A}$.

Taking into account equations~\eref{eq:Y} and~\eref{eq:iam_nonideal_device}, for each connection we have determined the short-circuit admittance matrix $\bi{Y} = \bi{A}^\textup{T}\bi{D}^{-1}\bar{\bi{Y}}_\textup{i}\bi{A}$ and the open-circuit impedance matrix $\bi{Z} = \bi{Y}^{-1}$. From $\bi{Z}$ it is then easy to determine the ideal values of the measured impedances and the relative errors due to the contact resistances. These results are shown in table~\ref{tab:single_device_connections}, and references are given for comparison.

\begin{table}
\caption{Examples of single device connections (see text for explanation).}\label{tab:single_device_connections}
\newsavebox{\matA}
\newsavebox{\matB}
\newsavebox{\matC}
\newsavebox{\errA}
\newsavebox{\errB}
\newsavebox{\errC}
\savebox{\matA}{%
\tiny$\left(\begin{array}{*{3}{@{\extracolsep{1pt}}c@{\extracolsep{1pt}}}}
1 &0 &0 \\
0 &1 &0 \\
0 &0 &0 \\
0 &0 &1 \\
1 &0 &0 \\
0 &1 &0 \\
0 &0 &0 \\
0 &0 &1 
\end{array}\right)$%
}
\savebox{\matB}{%
\tiny$\left(\begin{array}{*{5}{@{\extracolsep{1pt}}c@{\extracolsep{1pt}}}}
1 &0 &0 &0 &0\\
0 &1 &0 &0 &0\\
0 &0 &0 &0 &0\\
0 &0 &1 &0 &0\\
0 &0 &0 &0 &0\\
0 &0 &0 &1 &0\\
1 &0 &0 &0 &0\\
0 &0 &0 &0 &1 
\end{array}\right)$%
}
\savebox{\matC}{%
\tiny$\left(\begin{array}{*{5}{@{\extracolsep{1pt}}c@{\extracolsep{1pt}}}}
1 &0 &0 &0 &0\\
0 &1 &0 &0 &0\\
0 &0 &1 &0 &0\\
0 &0 &0 &1 &0\\
0 &0 &0 &0 &0\\
0 &0 &0 &0 &1\\
0 &0 &1 &0 &0\\
0 &0 &0 &1 &0 
\end{array}\right)$%
}
\savebox{\errA}{%
\footnotesize$\begin{array}{l}
\displaystyle\frac{(\epsilon_1-\epsilon_5)(\epsilon_2-\epsilon_6)}{16} \\[1ex]
\displaystyle +\frac{(\epsilon_3-\epsilon_7)(\epsilon_4-\epsilon_8)}{16}
\end{array}$%
}
\savebox{\errB}{%
\footnotesize$\displaystyle\frac{\epsilon_1\epsilon_7+\epsilon_3\epsilon_5}{4}$
}
\savebox{\errC}{%
\footnotesize$\displaystyle -\frac{(\epsilon_3+\epsilon_7)(\epsilon_4+\epsilon_8)}{16}$
}
\thispagestyle{empty}
\centering
\footnotesize
\begin{tabular}{*{2}{>{\centering}m{25mm}@{}}*{5}{c@{\extracolsep{1pt}}}}
\toprule
\parbox[c]{25mm}{\centering Connection schematics} & \centering Port voltages & $\bi{A}$ & \parbox[c]{20mm}{\centering Measured impedance} & \parbox[c]{10mm}{\centering Ideal value} & \parbox[c]{25mm}{\centering Relative error} & Ref. \\ \midrule
\includegraphics[scale=0.9,clip=]{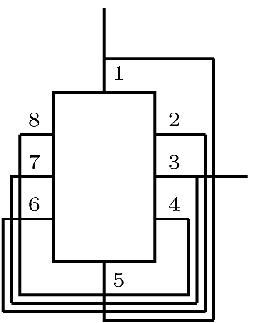} & \includegraphics[scale=0.9,clip=]{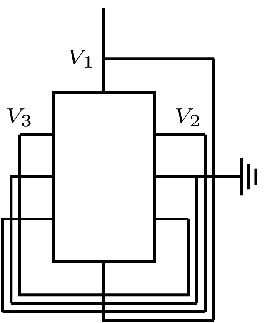} & \usebox{\matA} & $Z_{24,13}$ & $\displaystyle\frac{R_\textup{H}}{2}$ & \usebox{\errA} &\cite[eq.\ (17)]{Delahaye:1993} \\
\includegraphics[scale=0.9,clip=]{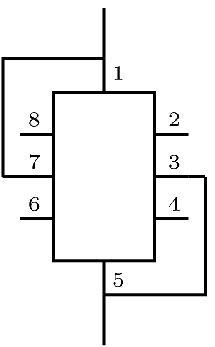} & \includegraphics[scale=0.9,clip=]{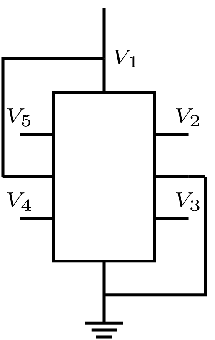} & \usebox{\matB} & $Z_{15,15}$ & $R_\textup{H}$ & \usebox{\errB} & \cite[eq.\ (7)]{Jeffery:1995} \\
\includegraphics[scale=0.9,clip=]{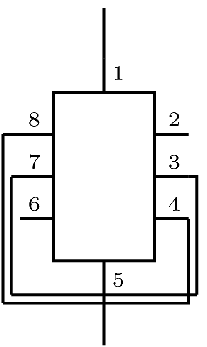} & \includegraphics[scale=0.9,clip=]{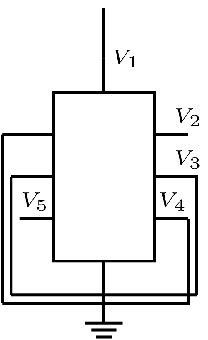} & \usebox{\matC} & $Z_{26,15}$ & $2R_\textup{H}$ & \usebox{\errC} &\cite[eq.\ (13)]{Delahaye:1993} \\
\bottomrule
\end{tabular}
\end{table}

\section{Examples: multiple device connections}
Two \emph{unconnected} multiterminal elements, $\mathcal{N}_\textup{A}$ with $n_\textup{A}$ terminals and $\mathcal{N}_\textup{B}$ with $n_\textup{B}$ terminals, can be considered as a single multiterminal element with $n_\textup{A}+n_\textup{B}$ terminals. Since there is no connection between the two elements, $\bar{y}_{ij} = 0$ unless terminals $i$ and $j$ belong to the same element. The indefinite admittance matrix of $\mathcal{N}_\textup{A}\cup\mathcal{N}_\textup{B}$ is thus a block diagonal matrix where the non-zero blocks are the indefinite admittance matrices of $\mathcal{N}_\textup{A}$ and $\mathcal{N}_\textup{B}$, respectively. In the case of two non-ideal QHE devices, the indefinite admittance matrix is then
\begin{equation}
\bar{\bi{Y}} = \left(\begin{array}{c|c}
\rule[-8pt]{0pt}{0pt}\bi{D}^{-1}_\textup{A}\bar{\bi{Y}}_\textup{i} & 0 \\ \hline 
0 & \rule{0pt}{15pt}\bi{D}^{-1}_\textup{B}\bar{\bi{Y}}_\textup{i}
\end{array}\right),
\end{equation}
where, for each device, $\bi{D}_\textup{A}$ and $\bi{D}_\textup{B}$ are given by \eref{eq:D}.

Given the matrix above, connections between two or more QHE devices can be treated following the same method described in section~\ref{sec:ideal_device_connections} for single-device connections. In particular, it is then easy to obtain the results determined by Delahaye in~\cite{Delahaye:1993} for multiple-series and multiple-parallel connections and to experiment with new configurations.

\section{Conclusions} 
We have described a method to model the electrical behaviour of one, or several, multiterminal QHE devices with complex external connections. The indefinite admittance matrix of the ideal unconnected device, a matrix representing the contact resistances and a matrix representing the external connections, all enter a simple matrix expression from which the two-terminal and four-terminal impedances of interest can be determined.

As examples of application, the method has been applied to connection schemes in the dc regime of practical interest for primary metrology, already analysed in other papers.

Formally, the method is valid also in the ac regime; however, the practical application requires to model the ac behaviour of connections, and parasitic parameters at the device level, and will be the subject of future investigations. 

\ack
The authors are grateful to Stefano Grivet-Talocia, Politecnico di Torino, for pointing out the existence of the indefinite admittance matrix, and are indebted with J{\"u}rgen Schurr, PTB Braunschweig (Germany): we performed together experimental testing of Fang's configurations and had many discussions on the subject. 

\section*{References}


\end{document}